\documentclass[11pt,twoside]{article}

\usepackage{asp2014}

\aspSuppressVolSlug
\resetcounters

\begin{document}

\title{Ten (or more!) reasons to register your software with the Astrophysics Source Code Library}

\author{Alice Allen$^{1,2}$ and Kimberly DuPrie$^{3,1}$}
\affil{$^1$Astrophysics Source Code Library (ASCL), Houghton, MI, USA; \email{aallen@ascl.net, kduprie@stsci.edu}}
\affil{$^2$Astronomy Department, University of Maryland, College Park, MD, USA}
\affil{$^3$Space Telescope Science Institute, Baltimore, MD, USA; \email{kduprie@stsci.edu}}

\paperauthor{Alice Allen}{aallen@ascl.net}{0000-0003-3477-2845}{ASCL/University of Maryland}{Astronomy Department}{College Park}{MD}{20742}{US}
\paperauthor{Kimberly DuPrie}{kduprie@stsci.edu}{}{Space Telescope Science Institute}{}{Baltimore}{MD}{21218}{US}



\begin{abstract}
This presentation covered the benefits of registering astronomy research software with the Astrophysics Source Code Library (ASCL, ascl.net), a free online registry for software used in astronomy research. Indexed by ADS and Clarivate’s Web of Science, the ASCL currently contains over 3,500 codes, and its entries have been cited over 17,000 times. Registering your code with the ASCL is easy with our online submissions system. Making your software available for examination shows confidence in your research and makes your research more transparent, reproducible, and falsifiable. ASCL registration allows your software to be cited on its own merits and provides a citation method that is trackable and accepted by all astronomy journals, and by journals such as \textit{Science} and \textit{Nature}. Adding your code to the ASCL also allows others to find your code more easily, as it can then be found not only in the ASCL itself, but also in ADS, Web of Science, and Google Scholar.
\end{abstract}



\section{Introduction}
The Astrophysics Source Code Library\footnote{\url{https://ascl.net}} (ASCL) is a free online repository and registry of source codes used in astronomy research that have appeared in or been submitted to peer-reviewed publications. Established at Michigan Technological University in 1999 by Robert Nemiroff and John Wallin \citep{Nemiroff1999}, its editors examine peer-reviewed papers for computational methods and create entries for the found software to the library. The ASCL also accepts submitted entries that meet its criteria.\footnote{\url{https://ascl.net/home/getwp/3593}} With over 3,500 entries, it covers a significant number of the astrophysics source codes used in peer-reviewed studies. 

This poster presentation encouraged researchers to register their software with the ASCL by sharing twelve reasons -- ten (or more!) -- they should consider doing so.

\section{The Ten (or more!) Reasons}
Though the ASCL was started primarily to share software and make research more transparent, it offers additional benefits to software authors and the astronomy community. 

\subsection{Get a unique identifier: an ASCL ID}
Software registered by the ASCL is assigned a unique identifier. This ASCL ID has the form ascl:\textit{yymm.xxx}, where \textit{yy} is the year, \textit{mm} is the month, and \textit{xxx} is an incremental number; each new month's incremental numbers start at .001. 

An ASCL ID is assigned after an entry is vetted and accepted by the ASCL. This unique identifier is recognized by NASA's Astrophysics Data System (ADS), the resolution service Identifiers.org,\footnote{\url{https://registry.identifiers.org/registry/ascl}} the Web of Science Data Citation Index,\footnote{\url{https://clarivate.com/academia-government/scientific-and-academic-research/research-discovery-and-referencing/web-of-science/data-citation-index/}} \textit{Science}, \textit{Nature}, and other journals, and by publishers including AAS Publications and Elsevier.

\subsection{It's free}
No fee; it's free! Free to submit code, free to find code. And it's free to download all of our entries in JSON, too, which you can do with the URL \href{https://ascl.net/code/json}{https://ascl.net/code/json}.

\subsection{Makes your code citable on its own merit}
Because the ASCL ID is unique to a code, it can be, and often is, used to cite computational methods in journal articles, proceedings, and theses. The ability to cite software is just as important as the ability to cite a scientific paper; the software citation provides credit to the author(s) of the code and improves the transparency and reproducibility of the research for which it was used. 

When a journal article has been written about a piece of software, that software paper is often used to cite the code. However, the FORCE11 Software Citation Guidelines \citep{smith_software_2016}, which many publishers support, state that ``\textit{Software citations should be accorded the same importance in the scholarly record as citations of other research products, such as publications and data}''. The FORCE11 guidelines also recognize the value of software papers, and say that ``\textit{a request from the software authors to cite a paper should typically be respected, and the paper cited \textbf{in addition to the software}.}'' (Emphasis ours.) The ASCL strongly supports this; when a code author has indicated how their code should be cited, that specification is shown in the ``Preferred citation method'' field of the ASCL entry for that software.

Many codes do not have a dedicated software paper, instead having been written to support research. Although the code author will sometimes request that the research paper be cited if the software is used, there is often no clear way to cite the code. The ASCL can help with that, as the ASCL ID can be used for the citation.\footnote{\url{https://ascl.net/wordpress/about-ascl/citing-ascl-code-entries/}}

\subsection{Indexed by ADS and Web of Science\//Allows others to find your code easily}
ASCL entries are indexed -- ingested -- by ADS and Web of Science. This not only broadens opportunity to learn about software registered by the ASCL from these indexing services, but because ADS's holdings also appear in Google Scholar, ASCL entries are found there as well. For example, the software PyBDSF (Python Blob Detection and Source Finder) has an ASCL entry,\footnote{\url{https://ascl.net/1502.007}} 
appears in ADS,\footnote{\url{https://ui.adsabs.harvard.edu/abs/2015ascl.soft02007M}} and also in Google Scholar,\footnote{\url{https://scholar.google.com/scholar?hl=en\&as\_sdt=0\%2C47\&q=PyBDSF}} making information about this package widely available not only to astronomers, but to those in other fields as well.

\subsection{Citations to ASCL entries are trackable by ADS}
\begin{figure}
    \centering
    \includegraphics[width=1\linewidth]{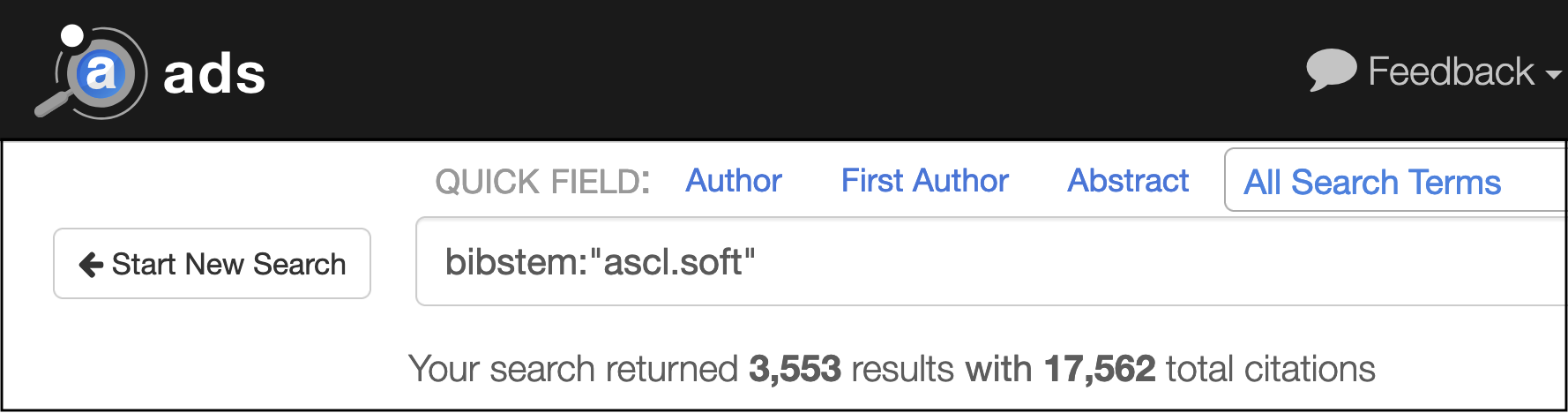}
    \caption{Citation count to ASCL entries as captured by ADS, retrieved 2024/12/15}
    \label{fig:1}
\end{figure}

Citations are trackable and counted by an indexer only if they appear in an article's bibliography \textit{and} the indexer doing the tracking can match bibliographic information to resources it contains. Because ADS ingests ASCL entries, ADS tracks and counts the citations to ASCL IDs that are found in the journals and other resources it indexes. As of the writing of this paper, ADS shows more than 17,000 citations to ASCL entries, as can be seen in Figure \ref{fig:1}, including one code with over 800 citations. More than 41\% of the codes in ASCL show citations in ADS by ASCL ID.

\subsection{Citations by ASCL ID accepted by over 250 journals including all major astronomy journals}
Citations by ASCL ID appear in more than 250 journals indexed by ADS, the top three of which are \textit{The Astrophysical Journal}, \textit{Monthly Notices of the Royal Astronomical Society}, and \textit{Astronomy \& Astrophysics}. There are, of course, more citations for codes listed in ASCL than indicated above, as not all codes are cited by ASCL ID.

ASCL IDs are used for citation even in journals outside astronomy \citep{Shamir}, as is seen for the UDAT code \citep{UDAT}. These citations may be captured by indexers such as Web of Science and Google Scholar.

\subsection{Associated links from ADS entries for papers to ASCL entries}
\begin{figure}
    \centering
    \includegraphics[width=0.25\linewidth]{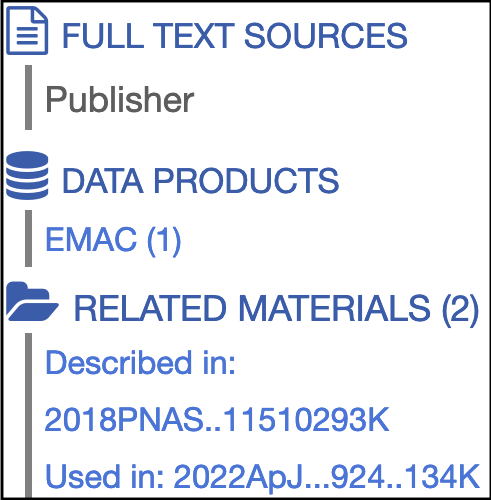}
    \caption{Related materials for the PyRADS software, as shown by ADS on 2024/12/15.}
    \label{fig:2}
\end{figure}

ADS uses links sent to it by the ASCL (and others) to link articles to software, software to data, data to articles, and so on, as shown in Figure \ref{fig:2}. These links reveal the different objects necessary to convey the full picture of a research project, making it easier to find them.

\subsection{Easy submissions form}
ASCL editors encourage software authors to submit their codes. Really! Please submit your software! ASCL has an easy-to-use online submissions form\footnote{\url{https://ascl.net/code/submit}} that is always available, with helpful information provided throughout the form. Submitted entries are displayed immediately for public viewing, and those that meet the resource's criteria\footnote{\url{https://ascl.net/home/getwp/3593}} are assigned an ASCL ID.

\subsection{Largest curated indexed resource for astronomy research codes in existence, with 3,500+ codes}
ASCL entries are curated; links to code sites are checked at least twice a week, with results posted on the ASCL's Dashboard,\footnote{\url{https://ascl.net/dashboard}} and records are examined for overall health regularly. The library covers all aspects of computational astrophysics. The resource has a full-text search and can serve as a discovery tool for software for one’s own research or by students to learn how common astronomical problems are solved computationally. Codes registered by the ASCL can also serve as benchmarks for one's own solutions to these problems.

\subsection{Shows confidence in your research}
Sharing your software, making your source code open for examination,  and registering it so others can find it shows your integrity and adherence to the scientific method.

\subsection{Makes astronomy research more falsifiable, transparent, reproducible ... better!}
Science is a ``show me'' endeavor. Computational methods are \textit{methods}, and science demands that research methods be revealed. This allows others not only to see how your research was conducted, but also enables others to verify your work and build upon it. 

\section{Conclusion}
A major goal of the ASCL is to strengthen and support published research findings by making the computational methods used in this research discoverable and available for examination. Registering your software makes research more transparent and also provides other benefits, including: 
\begin{itemize}
    \item making software citable in a trackable way, thereby providing a way for software authors to accrue citations for their computational contributions to the discipline;
    \item making it possible to easily find software that one may want to use, thereby making the field more efficient; and, 
    \item finding software one may want to examine to learn from, which is valuable not only for early career researchers, but also for experienced researchers looking to upgrade their skills and for teaching computational methods to students.
\end{itemize}

\acknowledgements The ASCL thanks the Heidelberg Institute of Theoretical Studies, University of Maryland College Park, and Michigan Technological University for their support, and research software authors everywhere for writing the codes that enable research.

\bibliography{tenreasons}  

\begin{thebibliography}{}
\expandafter\ifx\csname natexlab\endcsname\relax\def\natexlab#1{#1}\fi
\expandafter\ifx\csname url\endcsname\relax
  \def\url#1{\texttt{#1}}\fi
\expandafter\ifx\csname urlprefix\endcsname\relax\def\urlprefix{URL }\fi
\providecommand{\eprint}[2][]{\url{#2}}

\bibitem[{{Nemiroff} \& {Wallin}(1999)}]{Nemiroff1999}
{Nemiroff}, R.~J., \& {Wallin}, J.~F. 1999, in AAS Meeting Abstracts \#194, vol. 194 of American Astronomical Society Meeting Abstracts, 44.08

\bibitem[{{Shamir}(2017)}]{UDAT}
{Shamir}, L. 2017, {UDAT: A multi-purpose data analysis tool}, Astrophysics Source Code Library, record ascl:1704.002

\bibitem[{{Shamir}(2024)}]{Shamir}
--- 2024, AC, 2, 1628

\bibitem[{Smith et~al.(2016)Smith, Katz, Niemeyer, \& {FORCE11 Software Citation Working Group}}]{smith_software_2016}
Smith, A.~M., Katz, D.~S., Niemeyer, K.~D., \& {FORCE11 Software Citation Working Group} 2016, {PeerJ} Computer Science, 2:e86

\end{thebibliography}


\end{document}